\documentclass[twocolumn,showpacs,preprintnumbers,superscriptaddress,amsmath,amssymb]{revtex4}

\usepackage{graphicx,amssymb,mathrsfs,bm}

\newcommand{\be}{\begin{equation}}
\newcommand{\ee}{\end{equation}}
\newcommand{\bea}{\begin{eqnarray}}
\newcommand{\eea}{\end{eqnarray}}
\newcommand{\bsube}{\begin{subequations}}
\newcommand{\esube}{\end{subequations}}

\newcommand{\Eq}[1]{Eq.\,(\ref{#1})}

\newcommand{\la}{\langle}
\newcommand{\ra}{\rangle}
\newcommand{\ti}{\tilde}

\newcommand{\rmL}{{\rm L}}
\newcommand{\rmR}{{\rm R}}

\newcommand{\rmd}{{\rm d}}

\newcommand{\gam}{\gamma}
\newcommand{\Gam}{\Gamma}
\newcommand{\Dlt}{\Delta}

\newcommand{\GamL}{\Gamma_{\rm L}}
\newcommand{\GamR}{\Gamma_{\rm R}}
\newcommand{\dg}{\dagger}

\begin{document}
\draft

\title{Coulomb blockade double-dot Aharonov-Bohm interferometer:
       harmonic decomposition of the interference pattern }

\author{Feng Li}
\affiliation{State Key Laboratory for Superlattices and
Microstructures, Institute of Semiconductors,
Chinese Academy of Sciences, P.O.~Box 912, Beijing 100083, China}

\author{HuJun Jiao}
\affiliation{State Key Laboratory for Superlattices and
Microstructures, Institute of Semiconductors,
Chinese Academy of Sciences, P.O.~Box 912, Beijing 100083, China}

\author{Hui Wang}
\affiliation{Department of Mathematics and Physics,
China University of Petroleum, Beijing 102249, China}

\author{JunYan Luo}
\affiliation{Department of Chemistry, Hong Kong University
   of Science and Technology, Kowloon, Hong Kong }

\author{Xin-Qi Li}
\affiliation{State Key Laboratory for Superlattices and
Microstructures, Institute of Semiconductors,
Chinese Academy of Sciences, P.O.~Box 912, Beijing 100083, China}
\affiliation{ Department of Physics, Beijing Normal University,
Beijing 100875, China }

\date{\today}

\begin{abstract}
For the solid state double-dot interferometer,
the phase shifted interference pattern induced by the interplay of
inter-dot Coulomb correlation and multiple reflections
is analyzed by harmonic decomposition.
Unexpected result is uncovered, and is discussed
in connection with the which-path detection and electron loss.
\end{abstract}

\maketitle


{\it Introduction}.---
As an analogue of Young's double-slit interference \cite{Fey70},
a ring-like Aharonov-Bohm (AB) interferometer with a quantum dot
in one of the interfering paths is of interest for many fundamental
reasons and receives extensive studies
\cite{Hac01463,Yac954047,Buk98,Aha02}.
Very recently, an elegant study was further carried out for the
closed-loop setup, with particular focus on the multiple-reflection
induced inefficient which-path information
by a nearby charge detector \cite{Kang08}.

In this report we consider an alternative solid-state AB interferometer,
say, electron transport through parallel double dots (DD)
in Coulomb blockade regime, as schematically shown in Fig.\ 1.
Existing studies on this DD setup include the cotunneling interference
\cite{Los001035,Los-01,Kon013855,Sig06},
and two-loops (two fluxes) interference with the two dots
as an artificial molecule \cite{Hol0102,Jia02}.
In this work we restrict to the DD
\emph{without} tunnel-coupling between the dots,
which enables us to focus on
the interplay of Coulomb correlation and quantum interference.

At first sight, this electronic setup looks similar to
the double-slit interferometer \cite{Fey70}. Nevertheless,
the underlying solid-state specifics will result in such
novel behavior as non-analytic current switching
tuned by very weak magnetic field \cite{Li08}.
In present work, we extend our previous study \cite{Li08}
to more general situations for off-resonant DD levels,
in the presence of which-path detection and lossy channels.
We find the interference pattern to be {\it asymmetric}
with respect to the magnetic field, and attribute it
to the underlying Coulomb correlations.
Moreover, similar to Ref.\ \onlinecite{Kang08}, we present an
analysis of harmonic decomposition for the interference pattern.
We find that the phase shifts of the harmonic components of
current are linearly scaled with the order of harmonics,
and the high-order harmonics, in contrast to Ref.\ \onlinecite{Kang08},
will be dephased by the which-path detection.

\begin{figure}
\center
\includegraphics[scale=0.8]{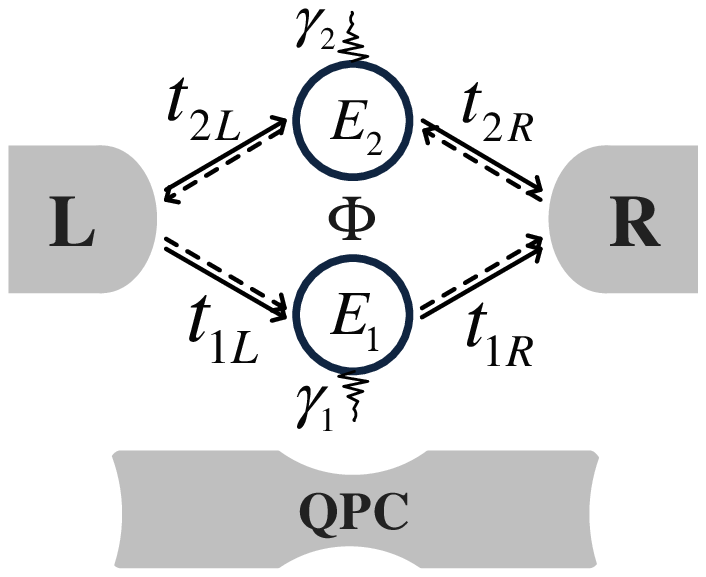}
\caption{
Schematic setup of a double-dot Aharonov-Bohm interferometer.
The solid-line trajectory is the usual first-order harmonic
interference, while the dashed-line trajectory represents
contribution to higher(second)-order harmonic interference.
To address dephasing and electron-loss effects,
a nearby quantum-point-contact (QPC) detector and lossy channels
to side-reservoirs (with rates $\gamma_{1(2)}$) are introduced. }
\end{figure}

{\it Model and Method}.---
Consider the double dots connected in parallel to two leads.
For simplicity we assume that in each dot
there is only one level, $E_{1(2)}$, involved in the transport.
Also, we neglect the spin degrees of freedom.
In the case of strong Coulomb blockade, the effect of spin
can be easily restored by doubling the tunneling rates
of each QD with the left lead.
The system is described by the following Hamiltonian
\begin{align}
H=H_0+H_T+\sum_{\mu =1,2} E_\mu d_\mu^\dagger d_\mu +Ud_1^\dagger
d_1 d_2^\dagger d_2\, . \label{a1}
\end{align}
Here the first term, $H_0=\sum_k [E_{kL}a_{kL}^\dagger a_{kL}
+E_{kR}a_{kR}^\dagger a_{kR}]$,
describes the leads and $H_T$ describes their coupling
to the dots,
\begin{align}
H_T=\sum_{\mu,k}\Big (t_{\mu L}d_\mu^\dagger a_{kL}
 +t_{\mu R}a_{kR}^\dagger d_\mu\Big )+{\rm H.c.}\, , \label{a2}
\end{align}
where $\mu=1,2$ and $a_{kL}^\dagger$ and~$a_{kR}^\dagger$ are the
creation operators for the electrons in the leads while
$d_{1,2}^\dagger$ are the creation operators for the DD. The last
term in Eq.~(\ref{a1}) describes the interdot repulsion. We assume
that there is no tunnel coupling between the dots and that the
couplings of the dots to the leads, $t_{\mu L(R)}$, are independent of energy.
In the absence of a magnetic field one can always choose
the gauge in such a way that all couplings are real.
In the presence of a magnetic flux $\Phi$, however, the tunneling
amplitudes between the dots and the leads are in general complex.
We write $t_{\mu L(R)}={\bar t}_{\mu L(R)}e^{i\phi_{\mu L(R)}}$,
where $\bar t_{\mu L(R)}$ is the coupling without the magnetic field.
The phases are constrained to satisfy
$\phi_{1L}+\phi_{1R}-\phi_{2L}-\phi_{2R}=\phi$, where $\phi\equiv
2\pi\Phi/\Phi_0$.

To account for dephasing effect, we introduce a which-path detection
by a nearby quantum point contact (QPC)\cite{Buk98},
with a model description as in Ref.\ \onlinecite{Wan0703745}.
To make contact with conventional double-slit interferometer,
we also introduce electron lossy channels.
Slightly differing from Ref.\ \onlinecite{Aha02},
instead of the semi-infinite tight binding chain introduced there,
we model the lossy channels by attaching each dot with an
electronic side-reservoir,
which is particularly suited in the master equation approach.
The side-reservoir model was originally proposed
by B\"uttiker in dealing with phase-breaking effect \cite{But8688},
i.e., electron would lose phase information
after entering the reservoir first, then returning back from it.
But here, we assume that the reservoir's Fermi level is much lower
than the dot energy.
As a result, electron only enters the reservoir
\emph{unidirectionally}, never coming back.

The transport properties of the above described system
can be conveniently studied by the number-resolved master equation
\cite{Gur96,Li05066803,Luo07085325}.
The central quantity of this approach is the number-conditioned
reduced state,
$\rho^{(n)}(t)$ of the double dots,
where $n$ is the electron number passed through the junction between
the DD and an assigned lead where number counting is performed.
Very usefully, $\rho^{(n)}(t)$ is related to the electron-number
distribution function, in terms of $P(n,t)={\rm Tr}[\rho^{(n)}(t)]$,
where the trace is over the DD states. From $P(n,t)$ the current
and its fluctuations can be readily obtained.
For current, for instance, it simply reads
$I(t)=ed \la n(t)\ra /dt$, where $\la n(t)\ra=\sum_n n P(n,t)$.
In practice, instead of directly solving $P(n,t)$,
much simpler equation-of-motion technique is available for the
calculation of current and current fluctuations \cite{Li05066803,Luo07085325}.

In large bias sequential tunnelling regime
and under inter-dot Coulomb blockade (i.e. the DD can be
occupied at most by one electron),
the Hilbert space of the DD is reduced
to $|0\ra \equiv |00\ra$, $|1\ra \equiv |10\ra$, and $|2\ra \equiv |01\ra$,
where $|10\ra$ means the upper dot occupied and the lower dot unoccupied,
and other states have similar interpretations.
Following Ref.\ \onlinecite{Luo07085325}, the ``$n$"-resolved master equation
in this basis can be straightforwardly carried out as

\begin{widetext}
\begin{subequations}
\bea
\dot{\rho}^{(n)}_{00}=
-2 \Gamma_L \rho_{00}^{(n)}
+\left(\gamma +\Gamma_R\right) \rho_{11}^{(n-1)}
+\left(\gamma +\Gamma_R\right) \rho_{22}^{(n-1)}
+e^{i \left(\phi_{\text{R1}}-\phi_{\text{R2}}\right)}
\Gamma_R \rho_{12}^{(n-1)}
+e^{i \left(\phi_{\text{R2}}-\phi_{\text{R1}}\right)}
\Gamma_R \rho_{21}^{(n-1)}
\eea
\bea
\dot{\rho}_{11}^{(n)}=
\Gamma_L \rho_{00}^{(n)}
-\left(\gamma +\Gamma_R\right) \rho_{11}^{(n)}
-\frac{1}{2} e^{i \left(\phi_{\text{R1}}-\phi_{\text{R2}}\right)}
\Gamma_R \rho_{12}^{(n)}
-\frac{1}{2} e^{i \left(\phi_{\text{R2}}-\phi_{\text{R1}}\right)}
\Gamma_R \rho_{21}^{(n)}
\eea
\bea
\dot{\rho}_{22}^{(n)}=
\Gamma_L \rho_{00}^{(n)}
-\left(\gamma +\Gamma_R\right) \rho_{22}^{(n)}
-\frac{1}{2} e^{i \left(\phi_{\text{R1}}-\phi_{\text{R2}}\right)}
\Gamma_R \rho_{12}^{(n)}
-\frac{1}{2} e^{i\left(\phi_{\text{R2}}-\phi_{\text{R1}}\right)}
\Gamma_R \rho_{21}^{(n)}
\eea
\bea
\dot{\rho}_{12}^{(n)}=
e^{i \left(\phi_{\text{L1}}-\phi_{\text{L2}}\right)}
\Gamma_L \rho_{00}^{(n)}
-\frac{1}{2} e^{i \left(\phi_{\text{R2}}-\phi_{\text{R1}}\right)}
\Gamma_R \rho_{11}^{(n)}
-\frac{1}{2} e^{i \left(\phi_{\text{R2}}-\phi_{\text{R1}}\right)}
\Gamma_R \rho_{22}^{(n)}
-\frac{1}{2}  \left(\gamma_d+2 \gamma +2 i \Delta +2 \Gamma_R\right)
\rho_{12}^{(n)}
\eea
\bea
\dot{\rho}_{21}^{(n)}=
e^{i \left(\phi_{\text{L2}}-\phi_{\text{L1}}\right)} \Gamma_L \rho_{00}^{(n)}
-\frac{1}{2} e^{i \left(\phi_{\text{R1}}-\phi_{\text{R2}}\right)}
\Gamma_R\rho_{11}^{(n)}
-\frac{1}{2} e^{i \left(\phi_{\text{R1}}-\phi_{\text{R2}}\right)}
\Gamma_R \rho_{22}^{(n)}
-\frac{1}{2}  \left(\gamma_d+2 \gamma -2 i \Delta +2 \Gamma_R\right)
\rho_{21}^{(n)}
\eea
\end{subequations}
\end{widetext}
Here $\dot{\rho}^{(n)}$ denotes the time derivative of the
``$n$"-conditional DD state.
Owing to the neglected spin degrees of freedom
in constructing the Hilbert space, as a compensation,
we have replaced $\Gamma_L$ with $2\Gamma_L$ to
equivalently restore the spin effect.
In the above equations,
$\Dlt=E_1-E_2$ is the disalignment of the DD levels.
$\Gamma_{L(R)}=2\pi D_{L(R)}|t_{L(R)}|^2$,
and $\gamma_{1(2)}=2\pi D_{1(2)}|t_{1(2)}|^2$,
are the respective coupling rates of the DD to the left and right leads,
as well as to the side reservoirs.
$D_{L(R)}$ and $D_{1(2)}$ are the density of states of the leads
and reservoirs, while $t_{L(R)}$ and $t_{1(2)}$ are the respective
tunneling amplitudes.
For simplicity, in this work we assume that $\gamma_1=\gamma_2=\gamma$,
and $\Gamma_L=\Gamma_R=\Gamma$ in the following numerical results.
Finally, $\gamma_d$ is the dephasing rate between the two dots,
caused by the which-path measurement of QPC \cite{Wan0703745}.


\begin{figure}
 \center
 \includegraphics[scale=0.6]{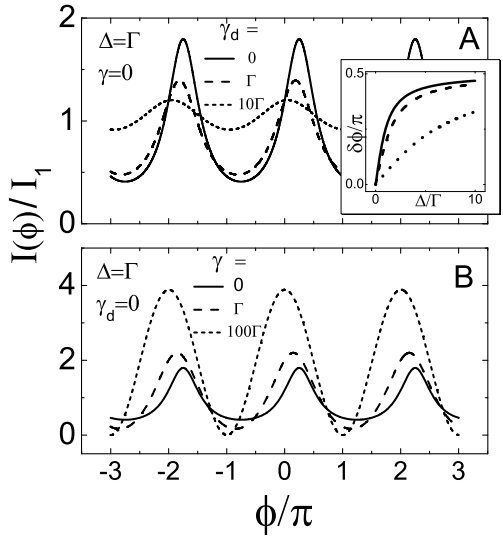}
 \caption{
(A) Phase shifted Aharonov-Bohm interference pattern, as a consequence
of the interplay of Coulomb correlation and quantum coherence.
Inset: shift of current peak \emph{versus} the level detuning $\Delta$.
Effect of dephasing between the two dots is demonstrated
by varying $\gamma_d$.
(B) Effect of electron loss. With increasing the lossy strength
$\gamma$, conventional double-slit interference pattern is eventually
restored. }
 \end{figure}

{\it Phase-Locking Breaking}.---
In the absence of electron loss, i.e., $\gamma=0$,
simple expression for the steady-state current is extractable:
 \be\label{I-2d-1}
 I=\left\{\frac{2(\gam_\rmd\!+\!2\Gam_\rmR)
 (1\!-\!\cos\phi)
 -4\Delta\sin\phi}{\gam_\rmd(\gam_\rmd+2\Gam_\rmR)
 +4\Delta^2}+\frac{1}{I_0}\right\}^{-1},
 \ee
where $I_0=4\Gam_\rmL\Gam_\rmR /(4\Gam_\rmL+\Gam_\rmR)$,
is the current in the absence of magnetic flux.
However, in the following we will use the current of
transport through a Coulomb-blockade single dot,
$I_1=2\GamL\GamR/(2\GamL+\GamR)$, to scale
the double-dot current, in order to highlight the interference features.

In the absence of dephasing,  $\gamma_d$=0,
from \Eq{I-2d-1} we have
$I= I_0 \Delta^2/\{\Delta^2+I_0[\Gamma_R(1-\cos\phi)-\Delta \sin\phi]\}$.
Then a remarkable switching effect follows this result:
as $\Delta\rightarrow 0$, $I=I_0$ for $\phi=2\pi n$,
while $I=0$ for any deviation of $\phi$ from these values.
Detailed interpretation for this novel behavior
using a $SU(2)$ transformation is referred to Ref.\ \onlinecite{Li08}.

If $\Delta\neq 0$, the Coulomb correlation is manifested in
an alternative way, see Fig.\ 2(A),
where the current peaks deviate from $\phi=2\pi n$.
For small $\Delta$, we find that the current is peaking
at $\phi \approx \Delta/\Gamma_R + 2\pi n$, and with a
magnitude $I_{\rm max}\approx (1/I_0-1/2\Gamma_R)^{-1}$.
Increasing $\Delta$, the current peak moves rightward.
This kind of peak shift implies nothing but the symmetry breaking
with respect to the magnetic-field inversion.
This behavior contrasts with the following statement \cite{But86}:
based on current conservation and time-reversal invariance,
the Onsager relation, say, the symmetry relation of transport
coefficients under inversion of magnetic field,
will lock the current peaks at $\phi=2\pi n$,
for \emph{any two-terminal linear transport}.
This is usually referred to as \emph{phase locking}.
Beyond linear response, however, phase locking
does not necessarily hold in general.

In Ref.\ \onlinecite{Kon013855}, for instance, for the AB setup
with a single dot embedded in one of the arms,
it was found that the phase locking is indeed broken
under finite bias voltage and
{\it only} in the presence of electron-electron interaction.
However, for the similar interacting DD system as considered here,
breaking of phase locking was not found under finite bias voltage,
in the case of $\Delta=0$ and within the
cotunneling transport mechanism \cite{Kon013855}.
Here, we find that the current is asymmetric
\emph{only} for $\Delta\neq 0$, while it is still symmetric for $\Delta=0$,
being somewhat in agreement with Ref.\ \onlinecite{Kon013855},
despite the different transport mechanism and dephasing involved here.

{\it Dephasing and Lossy Effects.}---
In Fig.\ 2(A), we also observe that
the current at $\phi=2n\pi$ is not affected by $\Delta$ and $\gamma_d$.
This feature implies that the current
in case of complete dephasing is the same as the one from constructive
interference (i.e. with $\phi=2\pi n$), being different from
the conventional double-slit interference \cite{Fey70}.
We attribute this feature to the closed boundary condition, under which
the multiple reflection plays essential role.
By introducing lossy channels, i.e., allowing electron loss
from the DD to surrounding environment,
in Fig.\ 2(B) we see that all these features disappear
and the conventional double-slit interference pattern
is restored by increasing the lossy rate $\gamma$.
The reason is that, as the dots become more and more open,
the side reservoirs would reduce the occupation probability on the dots,
thus make the Coulomb correlation and back-reflection less important.

\begin{figure}
\center
 \includegraphics[scale=0.66]{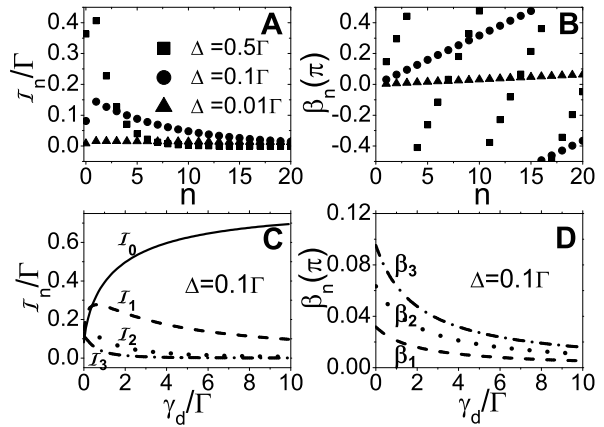}
 \caption{
Harmonic decomposition of the interference current $I(\phi)$.
(A) and (B): Amplitude ${\cal I}_n$ and phase shift $\beta_n$
of the $n$-th harmonic component, for different energy detuning $\Delta$.
In (B) an excellent linear-scaling relation for the phase shift,
say, $\beta_n=n\beta_1$, is observed.
(C) and (D): Effect of which-path detection on individual
harmonic components. In contrast to Ref.\ \onlinecite{Kang08},
here the higher order
harmonic components are also dephased (even more seriously).
Interestingly, from (D) we find that the scaling relation $\beta_n=n\beta_1$
still holds even in the presence of depahsing.}
 \end{figure}

{\it Harmonic Decomposition}.---
To highlight the effect of the closed nature
and Coulomb correlation,
we further expand Eq.\ (4) into Fourier series,
$I(\phi)={\cal I}_0+\sum_{n=1}^{\infty}{\cal I}_n\cos(n\phi+\beta_n)$.
Physically, the $n$-th order harmonic stems from trajectories
which have $n$-turns difference surrounding the magnetic flux
between the two interfering partial waves (see Fig.\ 1).
In Fig.\ 3 (A) and (B), we plot the amplitude ${\cal I}_n$
and phase shift $\beta_n$ for coherent case.
The higher order harmonics are caused by the multiple reflections
under closed boundary condition, while the phase shift is related to
Coulomb correlation between the two dots
(note that there is no phase shift for noninteracting DD).
Quite surprisingly, in Fig.\ 3(B) we observe that $\beta_n=n\beta_1$.
At this stage, unfortunately, we failed to find a satisfied
interpretation for this interesting result, so would remain it
for future study.

In Fig.\ 3 (C) and (D), dephasing effect is shown.
Note that, as explained in the model description, in this work
dephasing is modelled by a {\it which-path} detection.
Interestingly, we observe here that the which-path detection
also dephases the higher order harmonic components
of current, unlike in Ref.\ \onlinecite{Kang08},
where only the first-harmonic amplitude is strongly reduced
by the detection,
while the second one is almost insensitive to such detection.
We understand this discrepancy as follows.
Taking the second harmonic as an example,
the first partial wave has an amplitude $\propto t_1 e^{i\chi_1}$,
while the second partial wave
$\propto t_2 e^{i\chi_2}\cdot t^*_1 e^{i\tilde{\chi}_1}
\cdot t_2 e^{i\tilde{\chi}_2}$.
Here, we formally denote the transmission amplitude through
dot-1(2) by $t_{1(2)}$, which is in general complex,
e.g., containing the Aharanov-Bohm phase;
$\chi_{1(2)}$ and $\tilde{\chi}_{1(2)}$ are the random phases
caused by the charge detection when the electron passes through
dot-1(2) in the first- and second-order trajectories.
In Ref.\ \onlinecite{Kang08}, the argument leading to no (or weak) dephasing
of the second-harmonic amplitude
was based on the assumption that $\chi_1\simeq\tilde{\chi}_1$,
together with $\chi_2 = \tilde{\chi}_2 =0$
since there is no quantum dot in that arm.
Nevertheless, in the DD setup, $\chi_2 \neq \tilde{\chi}_2 \neq 0$
and $\chi_1\neq\tilde{\chi}_1$ in general, because of the time delay
between the two partial waves when arriving at the (same) dot.
Therefore, the nearby charge detection should dephase
the higher order interference trajectories, as shown in Fig.\ 3(C).
In this context, also of very interesting is the phase shift $\beta_n$,
still satisfying $\beta_n=n\beta_1$, even in the presence of dephasing.
Numerical result is shown in Fig.\ 3(D).

{\it Conclusion}.---
We have studied the electron transport through parallel quantum dots,
with highlight of the phase shift of the interference pattern
which is induced
by the interplay of inter-dot Coulomb correlation and quantum coherence.
In particular, a harmonic decomposition study for the pattern reveals
unexpected behavior of the phase shifts.
Dephasing effect is made connection with the
information gain of the individual harmonics
in the which-path detection,
while electron loss is investigated in relation to the conventional
double-slit interference.

\vspace{1cm}
{\it Acknowledgments.}---
This work was supported by the National Natural Science
Foundation of China under grants No.\ 60425412 and No.\ 90503013,
and the Major State Basic Research Project under grant No.2006CB921201.


\end{document}